\documentclass[twocolumn,english]{revtex4-1}
\usepackage[T1]{fontenc}
\usepackage[latin9]{inputenc}
\usepackage{amsbsy}
\usepackage{amstext}
\usepackage{amssymb}
\usepackage{graphicx}

\makeatletter
\usepackage{babel}

\usepackage{babel}

\usepackage{babel}

\usepackage{babel}

\usepackage{babel}

\makeatother

\usepackage{babel}
\begin{document}

\title{{\LARGE{}Anisotropic magnetic excitations and incipient Néel order
in Ba(Fe$_{1-x}$Mn$_{x}$)$_{2}$As$_{2}$}}

\author{Fernando A. Garcia$^{1}$, Oleh Ivashko$^{2}$ , Daniel E. McNally$^{3}$,
Lakshmi Das$^{2}$, Mario M. Piva$^{4}$, C. Adriano$^{4}$, Pascoal
G. Pagliuso$^{4}$, Johan Chang$^{2}$, Thorsten Schmitt$^{3}$, Claude
Monney$^{5}$ }

\affiliation{$^{1}$Instituto de Física, Universidade de São Paulo, São Paulo-SP,
05508-090, Brazil.}

\affiliation{$^{2}$Physik-Institut, Universitaet Zuerich, Winterthurerstrasse
190, 8057 Zurich, Switzerland.}

\affiliation{$^{3}$Photon Science Division, Swiss Light Source, Paul Scherrer
Institut, 5232 Villigen PSI, Switzerland}

\affiliation{$^{4}$Instituto de Física \textquotedblleft Gleb Wataghin\textquotedblright ,
UNICAMP, 13083-859, Campinas, SP, Brazil}

\affiliation{$^{5}$Physics Department, University of Fribourg, Ch. du Musée 3,
1700 Fribourg, Switzerland.}
\begin{abstract}
It is currently understood that high temperature superconductivity
(SC) in the transition metal $(M)$ substituted iron arsenides Ba(Fe$_{1-x}$$M$$_{x}$)$_{2}$As$_{2}$
is promoted by magnetic excitations with wave vectors $(\pi,0)$ or
$(0,\pi)$. It is known that while a small amount of Co substitution
leads to SC, the same does not occur for Mn for any value of $x$.
In this work, magnetic excitations in the iron arsenides Ba(Fe$_{1-x}$Mn$_{x}$)$_{2}$As$_{2}$
($x=0.0$, $0.007$, $0.009$, $0.08$) are investigated by means
of Resonant Inelastic X rays Scattering (RIXS) at the Fe $L_{3}$-edge,
for momentum transfer $\boldsymbol{q}$ along the high symmetry Brillouin
zone $(\pi,0)$ and $(\pi,\pi)$ directions. It is shown that with
increasing Mn content ($x$), the excitations become anisotropic both
in dispersion and lineshape. Both effects are detected even for small
values of $x$, evidencing a cooperative phenomenon between the Mn
impurities, that we ascribe to emerging Néel order of the Mn spins.
Moreover, for $x=0.08$, the excitations along $\boldsymbol{q}\parallel(\pi,0)$
are strongly damped and nearly non dispersive. This result suggests
that phases of arsenides containing local moments at the FeAs layers,
as in Mn or Cr substituted phases, do not support high temperature
SC due to absence of the appropriate magnetic excitations. 
\end{abstract}
\maketitle

\section{Introduction}

Iron based superconductors \citep{kamihara_iron-based_2008} encompass
a broad family of electronic correlated materials for which orbital
and spin excitations are believed to play a key role in determining
the system properties \citep{haule_coherenceincoherence_2009}. Of
special interest are the iron arsenides, of which BaFe$_{2}$As$_{2}$
is particularly well explored and the subject of much attention \citep{johnston_puzzle_2010,hosono_iron-based_2015}.

BaFe$_{2}$As$_{2}$ (BFA) undergoes an itinerant spin density wave
(SDW) phase transition at about $T_{N}=134$ K \citep{rotter_spin-density-wave_2008}.
For BFA, superconductivity (SC) is achieved by means of chemical substitution
on either Ba, Fe or As sites, which suppresses the SDW phase \citep{hosono_iron-based_2015}.
SC is observed with critical temperatures ($T_{SC}$) as high as $\sim20-30$
K in the case of transition metal substitution at the Fe site \citep{johnston_puzzle_2010,hosono_iron-based_2015}. 

It is known that the composition versus temperature ($x$ vs $T$)
phase diagram of the transition metal ($TM$) substituted Ba(Fe$_{1-x}TM_{x}$)$_{2}$As$_{2}$
systems presents an unexpected asymmetry, which concerns the fact
that SC is not observed for Cr and Mn substituted samples, that are
on the hole-doped side of the phase diagram \citep{thaler_physical_2011}.
The unique aspect related to Mn, or Cr substitution, is the presence
of strongly localized magnetic moments at the dopant site \citep{rosa_possible_2014,texier_mn_2012,lafuerza_evidence_2017},
contrasting with the moments observed for Co and K substituted phases,
which still present a high degree of itinerancy \citep{pelliciari_magnetic_2017}.
Indeed, magnetic moments within the FeAs layers seem to be detrimental
to SC, which in other cases has been shown to persist when magnetism
is formed between these planes \citep{jiang_superconductivity_2009-1,matusiak_doping_2011}.

The present work discusses magnetic excitations of Ba(Fe$_{1-x}$Mn$_{x}$)$_{2}$As$_{2}$
($x=0.0$, $0.007$, $0.009$, $0.08$) as probed by means of Resonant
Inelastic X-ray Scattering (RIXS), for momentum transfer $\boldsymbol{q}$
along the high symmetry Brillouin zone (BZ) $(\pi,0)$ and $(\pi,\pi)$
directions. Cooper pairing in the high temperature superconducting
phases in arsenides, likely a $s\pm$ (extended $s$-wave) phase,
is understood to be promoted by excitations with $(\pi,0)/(0,\pi)$
wave vectors, derived from the Fe electronic degrees of freedom \citep{korshunov_theory_2008,zhang_orbital_2009}.
Excitations with a $(\pi,\pi)$ wave vector would give rise to relatively
weaker superconducting phase, with a gap of $d$-wave symmetry \citep{korshunov_theory_2008,zhang_orbital_2009,fernandes_suppression_2013}.
Our experiment reveals that with increasing Mn content, the Fe derived
magnetic excitations become anisotropic, both in dispersion and lineshape.
In particular, the excitations along the $(\pi,0)$ direction are
strongly damped and nearly non dispersive. Therefore, it is found
that the system lacks the magnetic excitations that promote the most
efficient channel for inter orbital SC pairing \citep{zhang_orbital_2009}.
Thus, our findings contribute to understanding the absence of SC in
phases of arsenides containing Mn, in particular, or in phases containing
Cr, that also introduces strongly localized moments at the FeAs layers.
In addition, the Fe derived excitations are affected by small amounts
of Mn, evidencing a cooperative behavior of the Mn impurities, that
is ascribed to an emerging Néel order in the system.

The case of Mn substitution presents a particular interesting case
study \citep{kim_antiferromagnetic_2010,thaler_physical_2011,tucker_competition_2012,inosov_possible_2013}.
The observed behavior is involved and understood to be related to
the presence of strong $Q_{\text{Néel}}=(\pi,\pi)$ short range spin
fluctuations \citep{tucker_competition_2012,inosov_possible_2013}.
The decisive role of magnetic moments sitting in the Mn sites is clear
since, as suggested by X-ray Absorption (XAS) and Photoemission Spectroscopy
(PES), the Mn ions do not act as charge dopants \citep{suzuki_absence_2013}.

\begin{figure}
\begin{centering}
\includegraphics[scale=0.27]{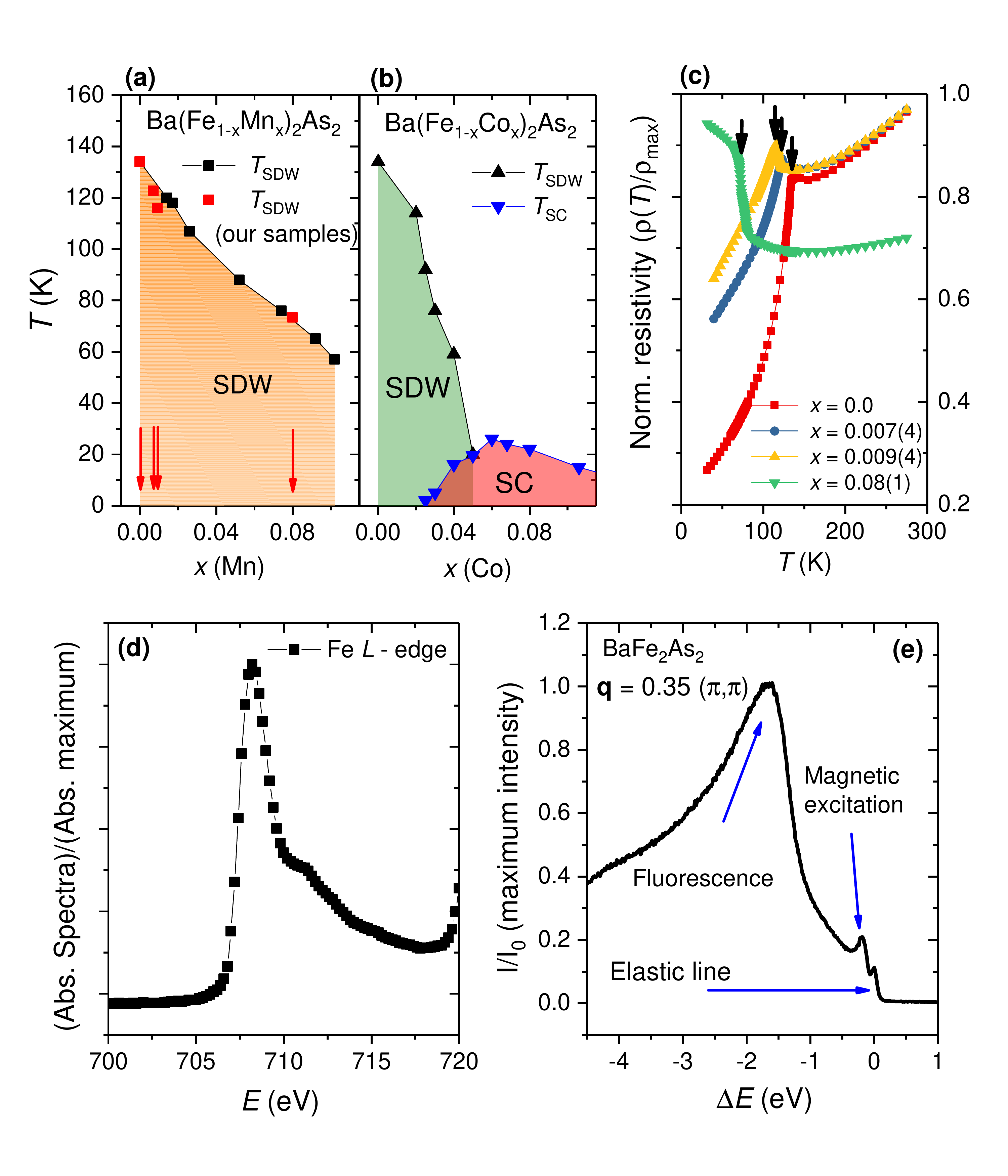} 
\par\end{centering}
\caption{(Color online) $x$ $vs.$ $T$ phase diagram for the $(a)$ Ba(Fe$_{1-x}$Mn$_{x}$)$_{2}$As$_{2}$
and $(b)$ Ba(Fe$_{1-x}$Co$_{x}$)$_{2}$As$_{2}$ transition metal
substituted iron arsenides (adapted from Ref. \citep{thaler_physical_2011}).
Red arrows in $(a)$ point to the compositions investigated in our
experiments. The transition temperatures were determined by the peaks,
or inflections, of the resistivity measurements as shown in $(c)$.
$(d)$ A typical XAS spectrum for the Fe $L_{3}$-edge is presented.
$(e)$ A representative RIXS spectrum is shown, wherein the main features
of the RIXS spectra collected in our experiments are pointed out.
\label{fig:phasepaper} }
\end{figure}

\section{Methods}

Our experiments were performed at the ADRESS beamline of the Swiss
Light Source at Paul Scherrer Institute \citep{ghiringhelli_saxes_nodate,strocov_high-resolution_2010}.
All RIXS spectra shown in this work have been acquired for an incident
photon energy tuned to the maximum of the Fe $L_{3}$- XAS edge and
for $\pi$-polarized incident photons. Each spectrum intensity has
been normalized to the maximum intensity of the fluorescence peak
$I_{0}$. The total energy resolution of the RIXS experiment was about
$95$ meV. The temperature was kept constant throughout the experiment
at $T=15$ K and a base pressure of $2\times10^{-10}$ mbar or better
was achieved. The $1$ - Fe magnetic (orthorhombic) unit cell is adopted
here for describing the orientation of the samples \citep{park_symmetry_2010}.
Ba(Fe$_{1-x}$Mn$_{x}$)$_{2}$As$_{2}$ ($x=0.0$, $0.007$, $0.009$,
$0.08$) single crystals were synthesized by the In-flux method as
described in Ref. \citep{garitezi_synthesis_2013}. The resistivity
of the samples was measured by employing a commercial Physical Properties
Measurements System (PPMS) from Quantum Design. 

\section{Results and Discussion}

In Figs. \ref{fig:phasepaper} $(a)$-$(b)$, $x$ vs. $T$ phase
diagrams for the Mn and Co substituted samples are presented (data
from Ref. \citep{thaler_physical_2011}). The red arrows and red squares
indicate (Fig. \ref{fig:phasepaper} $(a)$) the doping values at
which samples were investigated in our experiments. The transition
temperatures were determined by the peaks (or inflection points) on
the resistivity curves, as it is shown in Fig. \ref{fig:phasepaper}$(c)$.
By assuming that the SDW transition temperature is a qualitative measure
of the energy scale of the magnetic interactions between the Fe derived
itinerant spins in both systems, our Mn-substituted $x=0.08$ sample
and the $x\sim0.03$ Co- substituted system are to be compared. As
can be seen in Figs. \ref{fig:phasepaper} $(a)$-$(b)$, the systems
have similar SDW onset, yet only the Co-substituted compound is a
SC-SDW coexisting system.

Fig. \ref{fig:phasepaper} $(d)$-$(e)$ display, respectively, XAS
and RIXS spectra at the Fe $L_{3}$-edge of our pristine BaFe$_{2}$As$_{2}$
samples. The XAS spectrum obtained with $\pi-$polarized incident
photons is in good agreement with previously reported results of BaFe$_{2}$As$_{2}$
\citep{yang_evidence_2009,zhou_persistent_2013}. The RIXS spectrum
in Fig. \ref{fig:phasepaper}$(e)$ presents the main features observed
for a diverse range of compositions of iron arsenides: a broad fluorescence
line, reflecting the deexcitation of electrons from the valence bands
having a high density of states of Fe character, and the magnetic
excitation at low energy loss $\Delta E$. The later is well defined
and can be distinguished from the elastic peak at $\Delta E=0$. In
general, RIXS experiments of arsenides unveiled dispersive magnetic
excitations, that were found to have nearly isotropic dispersions
along the high symmetry BZ directions $(\pi,0)$ and $(\pi,\pi)$,
presenting characteristics that depend weakly on composition \citep{zhou_persistent_2013,pelliciari_intralayer_2016,pelliciari_local_2017,PelliSmFeAsO}.
Contrasting with this picture, the RIXS spectra presented in this
study display a different behavior as a function of composition.

\begin{figure*}
\begin{centering}
\includegraphics[scale=0.27]{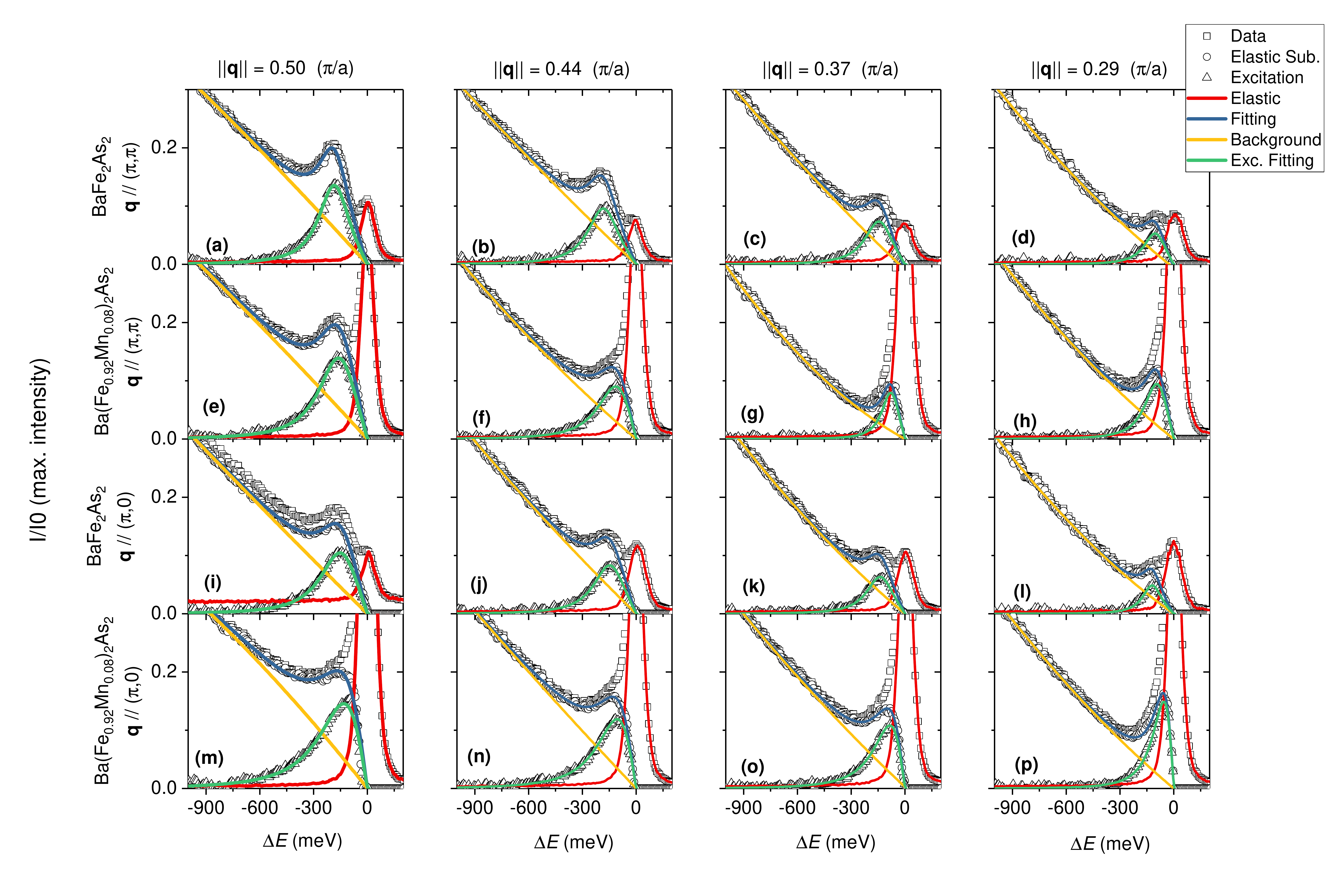} 
\par\end{centering}
\caption{(Color online) Survey of representative RIXS spectra and analysis
for the $x=0.0$ sample measured for $\boldsymbol{q}$ along the$(a)$-$(d)$
$(\pi,\pi)$ and $(i)$-$(l)$ $(\pi,0)$ directions, respectively,
and for the $x=0.08$ sample with $\boldsymbol{q}$ along the $(e)$-$(h)$
$(\pi,\pi)$ and $(m)$-$(p)$ $(\pi,0)$ directions, respectively.
Open symbols denote the experimental data (squares), the elastic subtracted
data (circles) and the magnon excitations obtained after the background
subtraction (triangles). The thick lines are fits of the spectra.
\label{fig:Figsfullspectra}}
\end{figure*}

In Figs. \ref{fig:Figsfullspectra}$(a)$-$(p)$ we show a representative
survey of our results, focusing the attention on the elastic line
and background subtraction, along with the magnetic excitation fitting.
The figures show data of the $x=0.0$ and $x=0.08$ samples, for distinct
$||\boldsymbol{q}||$ along the main directions of the BZ zone as
indicated. For the parent compound, even at low $||\boldsymbol{q}||$
(Figs. \ref{fig:Figsfullspectra}$(d)$ and $(l)$), the magnetic
excitation can be distinguished from the elastic peak at $\Delta E=0$.
Substitution by Mn, however, unavoidably introduces scattering centers
which increase the elastic line relative intensity, as observed for
all spectra obtained for the $x=0.08$ sample. Indeed, the elastic
lines of this sample overlap with the magnetic excitations. For $\boldsymbol{q}$
along $(\pi,\pi)$, the magnetic excitations are well defined for
the highest values of $||\boldsymbol{q}||$ and appear as weak bumps
for the lowest values of $||\boldsymbol{q}||$. For $\boldsymbol{q}$
along $(\pi,0)$, even for the highest value of $||\boldsymbol{q}||$
the magnetic excitations overlap with the elastic line, a situation
which calls for a careful procedure for its subtraction. 

We extract the elastic line contribution assuming that the RIXS signal
for $\Delta E>0$ is exclusively due to the elastic line and that
it is symmetric with respect to $\Delta E=0$ \citep{suzuki_probing_2018}.
We then mirror the data for $\Delta E>0$ and subtract it from the
respective RIXS signal, unveiling the underlining magnetic excitation
along with the fluorescence background. At low energy loss, the later
can be well described by a polynomial fitting (see yellow line in
Figs. \ref{fig:Figsfullspectra}$(a)$-$(p)$). The RIXS magnetic
excitation is proportional to the complex part of the dynamic susceptibility
$\chi''(\omega)$, that is described by a damped harmonic oscillator.
The associated expression reads: 

\[
\chi''(\omega)=\chi''_{0}\left(\frac{\Gamma\omega}{(\omega^{2}-\omega_{0}^{2})^{2}+\omega^{2}\Gamma^{2}}\right)
\]

\begin{equation}
=\frac{\chi''_{0}}{\omega_{\boldsymbol{q}}}\left(\frac{\Gamma/2}{(\omega-\omega_{\boldsymbol{q}})^{2}+(\Gamma/2)^{2}}-\frac{\Gamma/2}{(\omega+\omega_{\boldsymbol{q}})^{2}+(\Gamma/2)^{2}}\right)\label{eq:osharm}
\end{equation}

where $\chi''_{0}$ is a constant, $\omega_{0}$ is the excitation
bare frequency (without the damping effect), $\Gamma$ is the excitation
lifetime and $\omega_{\boldsymbol{q}}=\sqrt{\omega_{0}^{2}-(\Gamma/2)^{2}}$
is the propagation frequency of the excitation along the $\boldsymbol{q}$
direction. The final expression to fit the excitation is given by
$\chi''(\omega)(1+n_{B}(\omega))$ where $n_{B}(\omega)$ is the Bose-Einstein
distribution function ($n_{B}=(\exp(\hbar\omega/k_{\text{B}}T)-1)^{-1}$).
All the spectra of the parent compound are well described by the model
outlined above and the results are in good agreement with what was
found previously \citep{zhou_persistent_2013}. For the $x=0.08$
sample, all excitations exhibit large values of $\Gamma$, being strongly
damped. For $\boldsymbol{q}$ along $(\pi,0)$, in particular, the
magnetic excitatons are broadly spread in energy rendering $\Gamma\approx\omega_{0}$
which, in turn, suggests that the excitations along $(\pi,0)$ are
heavily damped or even overdamped for the $x=0.08$ sample.

\begin{figure}
\begin{centering}
\includegraphics[scale=0.27]{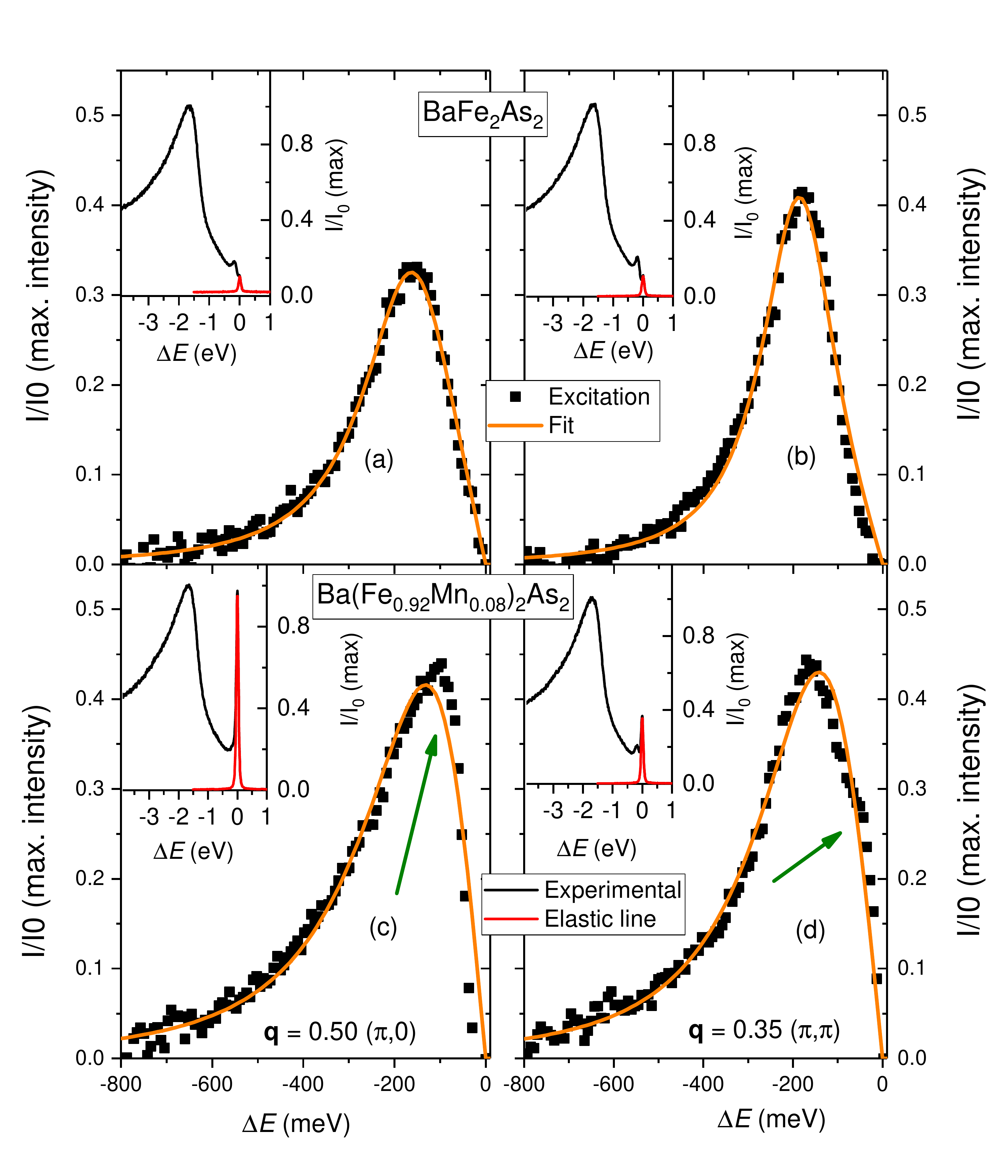} 
\par\end{centering}
\caption{(Color online) Excitations for the $x=0.0$ sample with $\boldsymbol{q}$
along the$(a)$ $(\pi,0)$ and (b) $(\pi,\pi)$ directions, respectively,
and excitations for the $x=0.08$ sample with $\boldsymbol{q}$ along
the $(c)$ $(\pi,0)$ and $(d)$ $(\pi,\pi)$ directions, respectively.
The thick lines are fits of the spectra to a damped harmonic oscillator.
The arrows in panels $(c)$-$(d)$ point to features in the lineshape
not captured by our model. The insets show the respective spectra
before the removal of both the elastic line and background. \label{fig:Figspectra}}
\end{figure}

A more detailed inspection of the obtained spectra is presented Figs.\ref{fig:Figspectra}$(a)$-$(d)$,
wherein the magnetic excitations for the Ba(Fe$_{1-x}$Mn$_{x}$)$_{2}$As$_{2}$
samples at the two extreme dopings used here, $x=0.0$ and $x=0.08$,
are presented. The excitations were measured at maximum accessible
(by RIXS) momentum transfer $\left\Vert \boldsymbol{q}\right\Vert $,
for $\boldsymbol{q}$ along the high symmetry BZ $(\pi,0)$ and $(\pi,\pi)$
directions. The respective raw spectra with the elastic-line fits
are shown in the insets. 

The contrast between the results for our samples at the two extreme
dopings, $x=0.0$ and $x=0.08$, is clear. For the parent compound,
the fit magnon excitation frequencies $\omega_{\boldsymbol{q}}$ along
both directions are about $\omega_{\boldsymbol{q}}\approx170$ meV,
with the excitation along $(\pi,0)$ being a little broader. In both
cases, the proposed model fits well to the experimental lineshape.
For the $x=0.08$ sample, one observes that for $\boldsymbol{q}$
along the $(\pi,\pi)$ direction the excitation peaks at $\approx-170$
meV and the fit frequency is $\omega_{(\pi,\pi)}\approx100(10)$ meV.
As for $\boldsymbol{q}$ along the $(\pi,0)$ direction, the excitation
peaks at about $-100$ meV and the excitation frequency is $\omega_{(\pi,0)}\approx60(15)$
meV \footnote{One must keep in mind that the line broadening usually shifts the
magnon excitation frequency to lower energies in comparison with the
peak position. Indeed in the damped harmonic oscillator model $\omega_{\boldsymbol{q}}=\sqrt{\omega_{0}^{2}-(\Gamma/2)^{2}}$
where $\omega_{0}$ is the bare frequency, closely corresponding to
the peak position and $\Gamma$ relates to the half maximum linewidth. }. It is noteworthy that for both directions the excitation are considerably
broader and there appears to be some extra spectral weight (indicated
by the arrows) close to $\Delta E=0$, that is not properly described
by the fitting function. For $\boldsymbol{q}$ along $(\pi,0)$, this
new feature overlaps strongly with the main excitation. Since for
$\boldsymbol{q}$ along $(\pi,\pi)$ the main excitation disperses
a little further, this extra component can be better distinguished.

These results are surprising since the magnetic excitations observed
by RIXS in doped iron arsenides did not show so far any considerable
evolution with composition \citep{zhou_persistent_2013,pelliciari_intralayer_2016,pelliciari_local_2017},
similarly to what has been found for hole-doped cuprates \citep{le_tacon_intense_2011,Dean_LSCO2013}.
Thus, for the first time, anisotropic magnetic excitations are observed
for iron arsenides by RIXS. It suggests, as well, that the excitations
are not properly described by a simple model of a damped harmonic
oscillator, which is unable to capture a very low energy feature close
to $\Delta E=0$, the intensity of which is increasing with Mn content.

\begin{figure}
\begin{centering}
\includegraphics[scale=0.27]{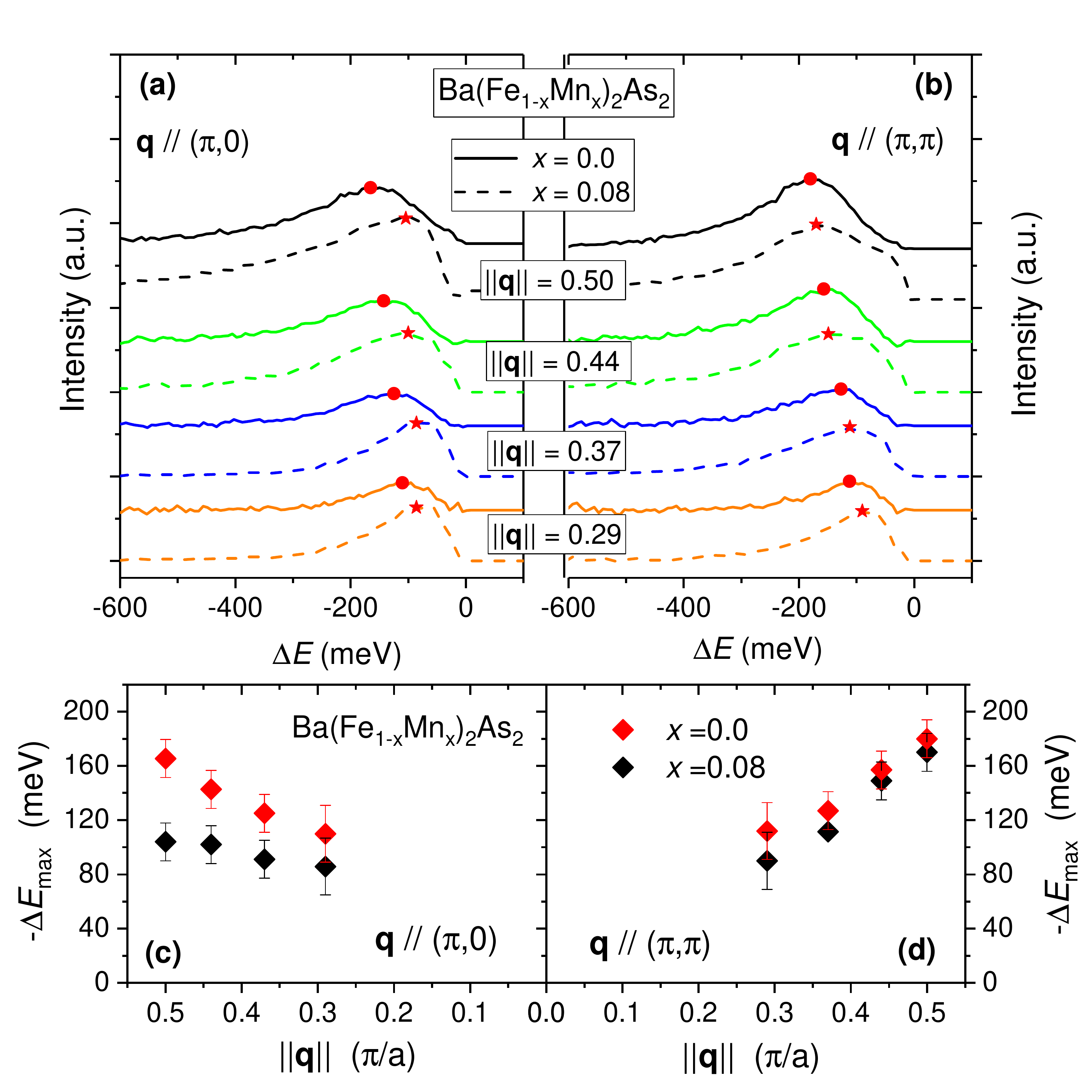} 
\par\end{centering}
\caption{(Color online) Magnetic excitations for momentum transfer $\boldsymbol{q}$
$(a)$ along the $(\pi,0)$ direction and $(b)$ along the $(\pi,\pi)$
direction ($x=0.0$ and $x=0.08$ samples) after removal of the fluorescence
and elastic contributions, as a function of momentum transfer $\left\Vert \boldsymbol{q}\right\Vert $.
Red dots and red stars mark the peak positions for the $x=0.0$ and
$x=0.08$ samples respectively. Energy loss (peak position), $-\Delta E_{\text{max}}$,
of the magnetic excitations as a function of $\left\Vert \boldsymbol{q}\right\Vert $
for $(c)$ $\boldsymbol{q}$ along the $(\pi,0)$ direction and ($d)$
$\boldsymbol{q}$ along the $(\pi,\pi)$ direction. \label{fig:spectraExcitations}}
\end{figure}

Therefore, to analyze the spectra obtained for other values of momentum
transfer, we adopt a phenomenological description, taking the peak
position at the maximum intensity, $\Delta E_{\text{max }}$, of the
low-energy excitation as a measurement of the excitation propagation
energy. The spectra and peak positions are shown in Fig. \ref{fig:spectraExcitations}$(a)$-$(b)$
for the $x=0.0$ and $x=0.08$ samples as a function of momentum transfer
$\left\Vert \boldsymbol{q}\right\Vert $ for the two directions being
considered ($\boldsymbol{q}$ along $(\pi,0)$, left panel; and $\boldsymbol{q}$
along $(\pi,\pi)$, right panel). Direct observation reveals that
while for $\boldsymbol{q}$ along $(\pi,\pi)$ the spectra of both
samples present a similar pattern as a function of $\left\Vert \boldsymbol{q}\right\Vert $
, for $\boldsymbol{q}$ along $(\pi,0)$ the excitation of the $x=0.08$
sample is nearly non-dispersive. The scenario drawn on the basis of
Fig. \ref{fig:spectraExcitations}$(a)$-$(b)$ is summarized in Fig.
\ref{fig:spectraExcitations}$(c)$-$(d)$, where the peak positions
are marked by red diamonds and black diamonds for the two cases, respectively.
The anisotropy of the excitation dispersion for the $x=0.08$ sample
is clear: while the low-energy excitation disperses up to $170$ meV
along $(\pi,\pi)$ in the $x=0.08$ sample, it hardly disperses from
$80$ to $100$ meV along $(\pi,0)$.

The observed anisotropy between the excitations along $(\pi,0)$ and
$(\pi,\pi)$ in Mn-doped iron arsenides, both in their dispersions
and lineshape, calls for an explanation. We start by discussing two
possible mechanisms addressing the specifics of the RIXS physics in
the arsenides. In short, these two mechanisms are (i) the increasing
contribution of electron-hole excitations at low-energy, overlapping
with a possibly weakened magnetic excitation or (ii) a magnetic excitation
overdamped by strong scattering with intralayer magnetic impurities.

(i) RIXS at the Fe $L_{3}$-edge probes both Fe derived spin and charge
excitations and both can in principle contribute to the low-energy
excitations close to the elastic line. It has been shown that electron-hole
excitations between bands close to the Fermi level can lead to dispersive
peaks in RIXS, for instance in the semimetal TiSe$_{2}$ \citep{monney_mapping_2012},
in the nickelate NdNiO$_{3}$ \citep{bisogni_ground-state_2016} or
in hole-doped cuprates \citep{kanasz-nagy_resonant_2016,bisogni_ground-state_2016,minola_crossover_2017}.
In the case of doped iron arsenides, doping-induced changes of the
electron bands at the BZ border could allow more electron-hole excitations
between bands near the Fermi level. This would mean an increase of
RIXS signal coming from electron-hole excitations upon doping along
$(\pi,0)$, since it is the vector connecting electron and hole pockets
at $(\pi,0)$ and at $\Gamma$ in iron arsenides. In the case of bands
crossing the Fermi level, one would expect to see in RIXS electron-hole
contributions already at zero energy loss. Thus, the small RIXS signal
at very low energy (see green arrows in Fig. \ref{fig:Figspectra})
could be due to some electron-hole excitations near the Fermi level.
However, it is rather weak and could be due to some imperfect subtraction
of the large elastic line at high $x$. We note that XAS and PES suggest
that the Mn ions do not act as charge dopants \citep{suzuki_absence_2013},
with the Mn $3d$ partial density of states distributed over an energy
range $2$-$13$ eV below the Fermi level.

(ii) A recent RIXS study of Eu-based iron arsenides has shown that
the magnetic excitations are mostly unaffected by the presence of
strong Eu$^{2+}$ $S=7/2$ local moments \citep{pelliciari_local_2017},
evidencing that the unique feature related to Mn substitution is the
appearance of strongly localized moments at the Mn sites \textit{inside}
the (Fe$_{1-x}$Mn$_{x}$)$_{2}$As$_{2}$ layers. The Mn impurities
introduce strong $Q_{\text{Néel}}=(\pi,\pi)$ short range spin fluctuations
in the system \citep{tucker_competition_2012,inosov_possible_2013},
which might have an indirect influence on the related magnetic excitations.

Indeed, here our main finding concerns the strong damping and the
softening of the magnetic excitation along the $(\pi,0)$ direction.
In the ordered SDW phase of pnictides, it is interpreted that the
magnons have predominant intra-orbital $xy\rightarrow xy$ character
\citep{kaneshita_spin_2011}. Orbitals with $xy$ character contribute
to bands whose extrema are centered around the BZ points $(\pi,0)$
and $(0,\pi)$, which are separated by a $(\pi,\pi)$ wave vector
\citep{zhang_orbital_2009,graser_near-degeneracy_2009}. Therefore,
$Q_{\text{Néel}}=(\pi,\pi)$ fluctuations are able to scatter electrons
with $xy$ character from distinct bands into one another, providing
a possible mechanism for understanding why the excitations along the
$(\pi,0)$ direction are strongly affected by Mn substitution. We
propose that $Q_{\text{Néel}}=(\pi,\pi)$ fluctuations lead to a transient
change of the occupation number of the electronic states of $xy$
orbital character at $(\pi,0)$ and $(0,\pi)$. We caution, however,
that RIXS at the Fe $L_{3}$-edge can reach in-plane momentum transfers
$\boldsymbol{q}$ up to $\sim0.5(\pi,0)$, meaning that we only probe
a fraction of the magnon dispersion.

We now turn our attention to the the relevance of our results to the
physics of the arsenides. We first call attention that one must keep
in mind that RIXS at the Fe $L_{3}$-edge probes Fe derived excitations.
Therefore, it is interesting that while K and Co substitution lead
both to SC, the RIXS probed excitations are largely unaffected by
the change of composition in these instances \citep{zhou_persistent_2013,pelliciari_intralayer_2016,pelliciari_magnetic_2017}.
In contrast, SC is not observed for phases with Mn, that is here shown
to be a strong scatter of the Fe derived excitations along the $(\pi,0)/(0,\pi)$
directions. Thus, the absence of superconductivity in pnictide samples
containing Mn spins can be understood in terms of electron scattering
by $Q_{\text{Néel}}=(\pi,\pi)$ fluctuations which suppress the high
temperature superconducting $s^{+-}$ phase, favoring the more fragile
$d$-wave superconductivity \citep{fernandes_suppression_2013}. 

\begin{figure}
\begin{centering}
\includegraphics[scale=0.27]{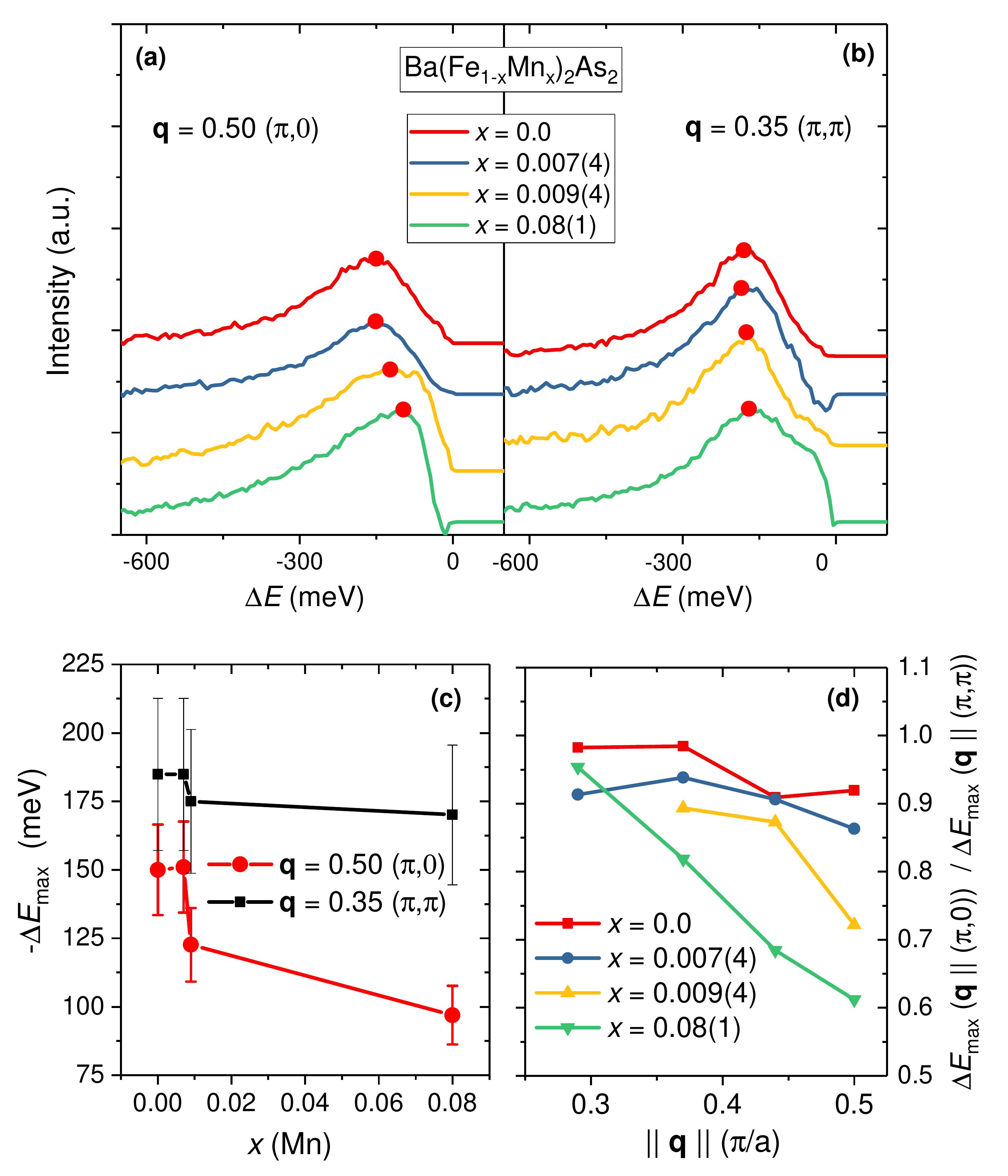} 
\par\end{centering}
\caption{(Color online) RIXS spectra for maximum momentum transfer, with $\boldsymbol{q}$
along the $(a)$ $(\pi,0)$ and $(b)$ $(\pi,\pi)$ directions, for
distinct $x$ (composition). The red dots mark the maximum of the
excitations. $(c)$ Energy loss, $-\Delta E_{\text{max}}$, of the
magnetic excitations as a function of $x$, for $\boldsymbol{q}$
along the $(\pi,0)$ and $(\pi,\pi)$ directions. ($d$ ) Evolution
of the anisotropy of the magnon modes as a function of momentum transfer
$||\boldsymbol{q}||$, as expressed by the ratio of $\Delta E_{\text{max}}(\boldsymbol{q})$
for $\boldsymbol{q}$ along the $(\pi,0)$ and $(\pi,\pi)$ direction,
for all investigated $x$. \label{fig:figexccitationmaximum}}
\end{figure}

The specifics of Mn substitution should also be addressed. Indeed,
the remarkable fact that small amounts of Mn impurities impact strongly
the Fe derived magnetic excitations was not yet discussed. This finding
suggests the kind of cooperative behavior of the Mn impurities as
discussed in Ref. \citep{gastiasoro_enhancement_2014} or a strong
coupling between the order parameter of the incipient Néel-type and
the SDW ordering wave vector $Q_{\text{SDW}}=(\pi,0)$ or $(0,\pi)$
as proposed in Ref. \citep{wang_impact_2014}. Such collective behavior
of Mn impurities is investigated in Fig. \ref{fig:figexccitationmaximum}$(a)$-$(d)$,
where an overview of the composition dependence of RIXS spectra is
presented. First we track in Fig. \ref{fig:figexccitationmaximum}$(a)$-$(c)$
the doping evolution of the magnon modes for maximum momentum transfer.
For all investigated samples, the modes along $(\pi,\pi)$ disperse
to about $170$ meV, while the modes along $(\pi,0)$ present a clear
trend toward softening starting from the $x=0.009$ sample (see also
Fig. \ref{fig:figexccitationmaximum}$(c)$). A more comparative scenario
is highlighted in Fig. \ref{fig:figexccitationmaximum}$(d)$. The
anisotropy in the dispersion of the magnon modes as a function of
$||\boldsymbol{q}||$ tends to disappear for small $||\boldsymbol{q}||$
and develops strongly with its increase, even for the $x=0.009$ sample.
This result illustrates the impact of the cooperative behavior of
magnetic disorder on the Fe-derived magnetic excitations due to the
growing Mn impurity density that is related to an incipient Néel order.

\section{Summary and Outlook}

Magnetic excitations in Mn substituted iron arsenides were investigated
by means of RIXS at the Fe $L_{3}$-edge. The Fe derived excitations
were shown to be composition and momentum dependent, in contrast to
previous RIXS experiments of the arsenides. Our pnictide sample with
the highest Mn doping ($x=0.08$) display magnetic excitations along
the directions $(\pi,0)$ and $(0,\pi)$ that are strongly damped
and softened. It suggests that the absence of SC in phases of arsenides
with Mn can thus be naturally explained by the lack of the $(\pi,0)$
and $(0,\pi)$ fluctuations that promote the most efficient Cooper
pairing channel. The cooperative behavior of Mn impurities was also
highlighted. It was shown that small amounts of Mn have a large impact
on the Fe derived excitations. 

We propose that the intraorbital character of the Fe derived magnons
is key to understand why the introduction of $Q_{\text{Néel}}=(\pi,\pi)$
fluctuations, derived from the Mn local spins, are able to scatter
the magnons along the $(\pi,0)$ and $(0,\pi)$ directions. Certainly,
more theoretical and experimental work is needed to better understand
this phenomenon. In this regard, the formal valence of Mn in Ba(Fe$_{1-x}$Mn$_{x}$)$_{2}$As$_{2}$
is $2+$, thus Mn impurities carry a large $S=5/2$ spin. It remains
to be investigated if our results are due to the specifics of this
spin configuration, or if it is a general feature concerning the presence
of strongly localized moments at the FeAs layers. In this line of
thinking, it is interesting to investigate phases containing Cr, which
do not present SC as well. The investigation of Cu containing samples
is also invited. Although Cu leads to a SC phase \citep{hosono_iron-based_2015}
the critical temperature is low and that could be due to the scattering
of $(\pi,0)/(0,\pi)$ fluctuations. In this case, one expects $d$-wave
SC in Cu containing samples. 
\begin{acknowledgments}
F.A. Garcia would like to acknowledge FAPESP (Grant No 2016.22471-3)
for financial support and the University of Zurich for financial support
and hospitality. D.M.N. and T.S. acknowledge support by the Swiss
National Science Foundation through the NCCR MARVEL. O.I., L.D., and
J.C. acknowledge support from the Swiss National Science Foundation
through the SINERGIA network Mott Physics beyond the Heisenberg Model
and grant number BSSGI0\_155873. L.D. is partially funded by a Swiss
Goverment PhD excellence scholarship. M.M.P., C. A. and P.G.P. acknowledge
FAPESP (Grants No 2015/15665-3, No. 2017/25269-3, and No. 2017/10581-1).
C.M. acknowledges the support by the SNSF grant No. PZ00P2\_154867
and PP00P2\textbackslash\_170597. The experiments were performed
at the ADRESS beamline of the Swiss Light Source at the Paul Scherrer
Institute. The authors acknowledge Rafael Fernandes for an illuminating
discussion on the magnetic and electronic properties of the Iron Arsenides. 
\end{acknowledgments}

\bibliographystyle{apsrev4-1}
\bibliography{GarciaFAMonneyC_RIXSBFMA}

\end{document}